# Extended depth of field of diffraction limited imaging system using spatial light modulator based intensity compensated polarization coded aperture

VIPIN TIWARI[1], NANDAN S. BISHT[1, 2*]

[1]*Applied Optics & Spectroscopy Laboratory, Department of Physics, KU, SSJ campus Almora-263601, Uttarakhand (India)*
[2]*Soban Singh Jeena (SSJ) University, Almora-263601, Uttarakhand (India)*
*Corresponding author: bisht.nandan@gmail.com*

**ABSTRACT**

**Reducing the aperture size is a conventional technique to obtain enhanced image resolution in optics but it is obscured by depleting illumination. Polarization coded apertures (PCAs) can be employed to circumvent this critical artifact. We experimentally demonstrate intensity compensated polarization encrypted apertures, which are designed using the polarization modulation characteristics of LC-SLM. PCAs are not limited by the aperture size and hence far-field point spread function (PSF) can be more conveniently recorded using these PCAs. We experimentally validate that Depth of field (DOF) of a diffraction limited lens and axial intensity of binary Fresnel zone plate (BFZP) is enhanced using PCAs with nominal intensity loss.**

## INTRODUCTION

Coded apertures are specifically designed structures, which can be constructed from the rearrangement of spatially varying (transparent and opaque) regions aiming to optimize the imaging characteristics of an optical system[1]. The imaging characteristics of an optical system are primarily described in terms of its Depth of Field (DOF), Field of View (FOV), axial intensity, spatial, and temporal resolution, etc [1–3]. With the emerging applications of imaging techniques in various multidisciplinary research dimensions, it is required to obtain improved imaging characteristics of an optical system. In general, conventional imaging systems having only source and detector suffer from various artifacts like limited DOF, narrower spatial and angular resolution, low signal to noise ratio (SNR), etc [3–5]. In practical imaging applications, the image of a point source is not necessarily results as a point in the detector plane. In fact, it is a patch of light distributed over a finite region in the image plane. This is represented by a crucial characteristic of an optical imaging system, known as the point spread function (PSF)[5, 6]. The correlation of this PSF with object provides the crucial image characteristics of the imaging system. Previous studies suggest that the introducing coded apertures between source and detector plane provide a solution to overcome these undesired artifacts in image acquisition[1, 7–11]. The primary function of coded apertures is to provide optimal modulation of the incident optical flux without scarifying its inherent characteristics such that the encoded image data at the image plane can be uniquely decoded into original source distribution using image reconstruction algorithms. One of the robust objectives of conventional imaging techniques is to obtain the extended DOF for an imaging system without reducing its resolution and light-gathering capabilities. DOF of diffraction limited lens is inversely proportional to the area of the aperture[8, 12]. Therefore, reducing aperture size to obtain extended DOF is a common practice in conventional imaging applications. A variety of apodizations i.e. geometric apertures[9], programmable apertures[4, 13] have been also incorporated in the recent past to achieve the extended depth of field (DOF) for an imaging system. Few studies report that the imaging systems encoded with an annular aperture i.e. ring-shaped apertures can be optimized to obtain larger DOF and enhanced axial intensity as compared to circular apertures[8, 14–16]. Further, Fresnel zone plates with axial separation by Rayleigh limit of resolution can also be employed to extend the depth of field of the imaging system[16–19]. The binary zone plates, i.e., Beynon zone plate [20, 21], Gabor zone plate[21], etc.) are utilized to improve the imaging resolution as well. However, these apertures (binary zone plates) diminish a half portion of the overall light throughput of the system and therefore hinder the utility of these apertures, especially in low light imaging applications. In further developments for coded aperture techniques, frequency response characteristics of a perfect lens partially masked by a retarder have been discussed [22]. Besides, the programmable liquid crystal (LC) devices are employed in wavefront–coded imaging systems as well[5, 23]. Phase masks have been designed using SLM, which resulted in narrower PSF and higher DOF for imaging system. Recent endeavors on CA imaging techniques involve few modern apertures, i.e., Computer Generated Holograms (CGH)[24, 25], spatial filtering[26], coded aperture correlation holography (COACH)[2], and Fresnel incoherent correlation holography (FINCH)[13, 27]. Among all these approaches, Polarization coded apertures can have plethora of advantages over other techniques as these apertures can be easily optimized using programmable optical devices i.e. Spatial Light Modulator (SLM) accordingly and thus provides dynamic applications of respective imaging systems.

Polarization coded apertures have the potential to optimize the imaging characteristics of an imaging system without compromising total light throughput. These apertures account for the polarization of light in order to encode the desired pattern in conventional apertures. In addition, polarization coded masks can be employed to achieve super-resolution imaging as well[28]. One of the major development in this direction involves an analytical study of a polarization coded annular aperture, proposed by Chi et al.[12]. They investigated the imaging properties of a diffraction limited lens under different defocus conditions. However, it is only limited to the theoretical consideration and does not validate their outcomes in an experimental framework. Recently, liquid crystal (LC) based programmable optical devices i.e. SLMs are emerging as key components in advanced holographic techniques (COACH, FINCH, etc.) and corresponding imaging systems can yield exceptional imaging characteristics of the respective imaging system. In a recent development on COACH, a virtual point spread function (VPSF) is introduced to enhance the image resolution of an imaging system using wavefront modulation[29]. However, their proposed experimental design requires two independent light sources (coherent source and LED) and therefore constrains the stability and compactness of the system. In this paper, we experimentally demonstrate two different types of Polarization coded apertures (PCA), which enable to extend the DOF of a diffraction limited imaging system significantly without scarifying the total light throughput. The PCAs are experimentally constructed using the polarization modulation characteristics of a reflective LC-SLM. In particular, a ring-shaped (annular) aperture is imported in SLM and then it is coded with orthogonal polarization states of light by applying different grayscales (pixel voltage) of LC-SLM. The proposed experimental design is robust and compact as it requires fundamental optical components (light source, SLM, and, a beam splitter).

**METHODOOGY**

The schematic of the proposed experimental design is depicted in fig. 1. A linearly polarized (x-polarized) light beam coming from a coherent light source i.e. He-Ne laser (wavelength λ= 632.9 nm) is collimated by a converging lens of focal length 15 cm. This collimated beam (plane wave) is exposed to a reflective LC-SLM (Holoeye, LC-R720). The PCA is projected into the SLM screen and reflected light is allowed to pass through a bi-convex lens ($L_2$=15 cm), which is placed adjacent to the PCA. Corresponding PSFs of the emergent Gaussian beam are recorded at the back focal plane of $L_2$ using a Charged coupled device (Procilica, GX-2750). Inset of Fig. illustrates the schematic for PCA design. A circular aperture of radius 'b' is separated into two regions of the equal-area i.e. the central circular region with a radius of 'a' and the outer circular region with b =√2a. The inner central region of radius 'a' is illuminated with incident SOP (x-polarized) of light while the reflected field from the outer annular region is rotated by $90^0$ with the help of the LC-SLM. It is reported in our previous work that when a pixel voltage (Gray Scale (GS)) of 180 is provided to the SLM, it rotates the state of polarization (SOP) of input light by $90^0$[30]. The PCA is designed using an annular aperture with two orthogonal polarization states for its two regions (inner region and outer ring) based on the specific gray scale dependent phase modulation characteristics of the SLM (fig.(1(b)). Therefore, two orthogonal field components are encoded in two regions of coded aperture i.e., horizontally polarized (x-polarized) and vertically polarized (y-polarized) in the inner circular region of radius 'a' and outer circular region of radius b =√2a respectively.

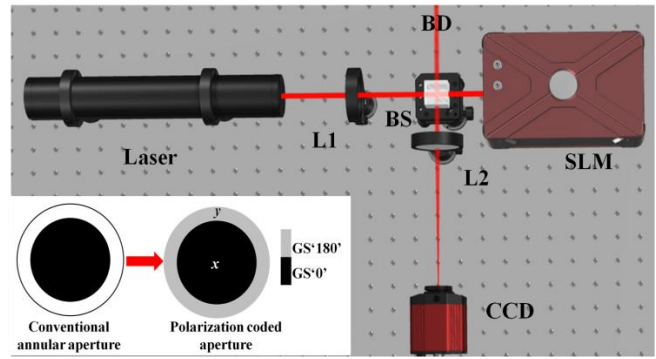

Fig.1 Experimental setup for the PCA based imaging system. Inset represents the schematic of the PCA design (grayscale mapping on SLM). [$L_1$, and $L_2$: lenses, BS: beam splitter, CCD: charge coupled device, SLM: spatial light modulator, BD: beam dumped].

To characterize the resolution of the PCA enabled imaging system, DOF of the emergent Gaussian beam (which is two times the $Z_R$) is measured by determining the Rayleigh range ($Z_R$) of the output beam at the back focal plane of lens $L_2$. The Gaussian beam profiles corresponding to the apertures with different numerical apertures (NA) are illustrated in fig. 2. Fig. 2 (a-c) represent the full aperture [radius (b) =2.5 mm], centre aperture with reduced area [radius (b/√2)=1.76 mm], and PCA [radius (b-(b/√2))=0.74 mm]. Corresponding beam profiles are shown in fig. 2 (d-f) respectively. It is evident that reducing the area of full aperture to half, enhances (doubles) the Rayleigh range ($Z_R$) but hinders the intensity of input light. In fact, total input intensity is reduced to its one-fourth while reducing the aperture size (area) to half of the full aperture (fig. (2(d,f))). On the other hand, the $Z_R$ corresponding to PCA is closest to the centre aperture whereas the intensity for PCA is two times of the centre aperture. The IPSF of the diffraction limited system for the central aperture (radius 'a') is given as [12]

$$I_C(\rho) = \left| \int_0^a r' J_0\left(2\pi r' \frac{\rho}{\lambda z}\right) exp\left[i \frac{2\pi W}{\lambda}\left(\frac{r'}{b}\right)^2\right] dr' \right|^2 \quad (1)$$

Similarly, the IPSF of the diffraction limited system for the outer ring aperture can be given as

$$I_R(\rho) = \left| \int_a^b r' J_0\left(2\pi r' \frac{\rho}{\lambda z}\right) exp\left[i \frac{2\pi W}{\lambda}\left(\frac{r'}{b}\right)^2\right] dr' \right|^2 \quad (2)$$

The PCA is designed by combining two regions, i.e. central and outer ring. Therefore, The IPSF of PCA can be defined as

$$I_{PCA}(\rho) = I_C(\rho) + I_R(\rho) \quad (3)$$

$$I_{PCA}(\rho) = \left| \int_0^a r' J_0\left(2\pi r' \frac{\rho}{\lambda z}\right) exp\left[i \frac{2\pi W}{\lambda}\left(\frac{r'}{b}\right)^2\right] dr' \right|^2 + \left| \int_a^b r' J_0\left(2\pi r' \frac{\rho}{\lambda z}\right) exp\left[i \frac{2\pi W}{\lambda}\left(\frac{r'}{b}\right)^2\right] dr' \right|^2 \quad (4)$$

Here, '$\rho$' is the radial coordinate of the IPSF, '$\lambda$' is the wavelength of incident light, 'W' is the defocus coefficient in the unit of the number of wavelengths. '$J_0$' denotes the Bessel function of first type order zero and 'z' is the propagation distance i.e., the distance between PCA and image plane.

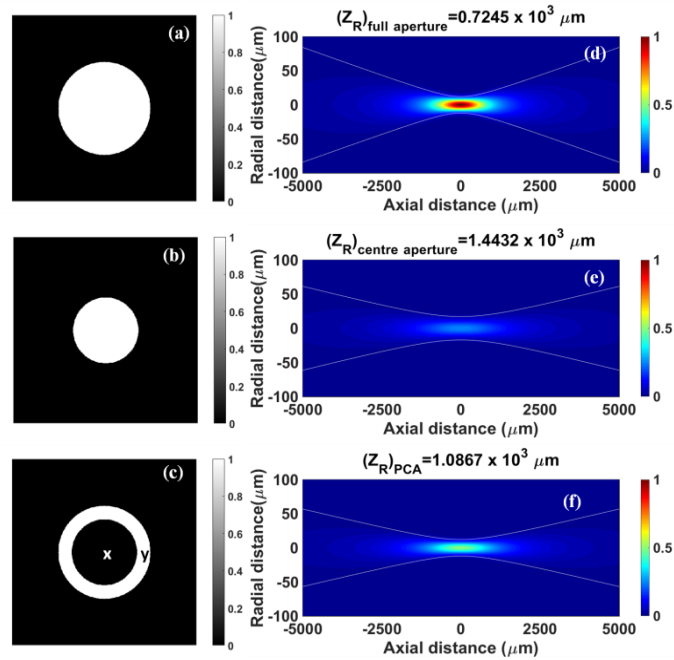

Fig. 2. Computer generated apertures (CGAs) and corresponding beam profiles. full aperture (a, d), centre aperture (b, e), and PCA (c, f) respectively. [PCA: polarization coded aperture, $Z_R$: Rayleigh range].

**RESULTS & DISCUSSION**

In order to validate the utility of the PCA, the PSFs of the diffraction limited system are recorded at the back focal point of lens $L_2$. Fig. 3 (a-c) represents the simulated PSFs for full aperture (a), centre aperture (b), and PCA (c). Corresponding experimentally recorded PSFs are shown in figs. 3 (d-f). It is observed that at the focused position of lens $L_2$, the total intensity is reduced to one-fourth for central aperture and a half for PCA. These observations indicate that the PCA enhances the DOF of the respective imaging system with minimal intensity loss in contrast to the conventional apertures with reduced size. A quantitative comparison for the IPSF of apertures is illustrated in fig. 4. Fig. 4 (a), (b) illustrate the simulated and recorded (experimental) IPSFs corresponding to the aforementioned apertures at the back focal plane of lens $L_2$.

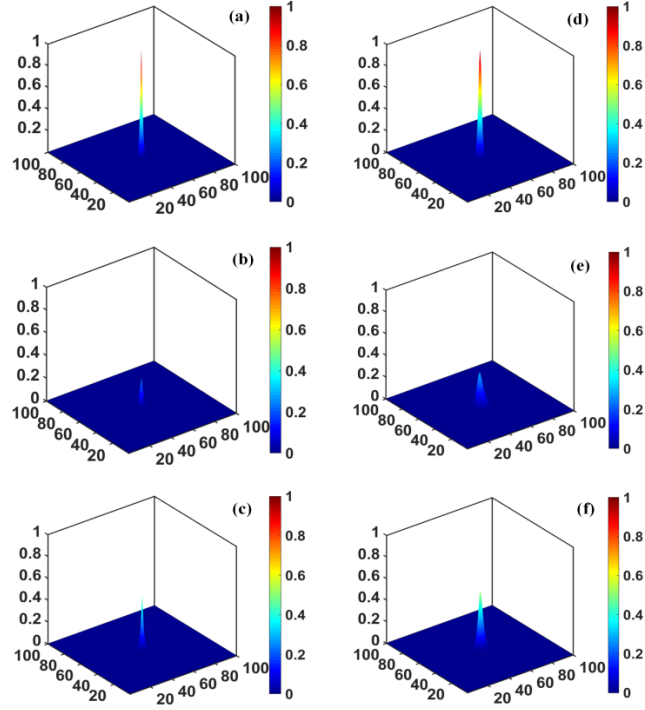

Fig. 3 PSF profile for apertures (a-c) simulated PSFs (d-f) experimentally recorded PSFs of diffraction limited lens (f=15 cm) corresponding to the full aperture (a, d), centre aperture (b, e), and PCA (c, f) respectively.

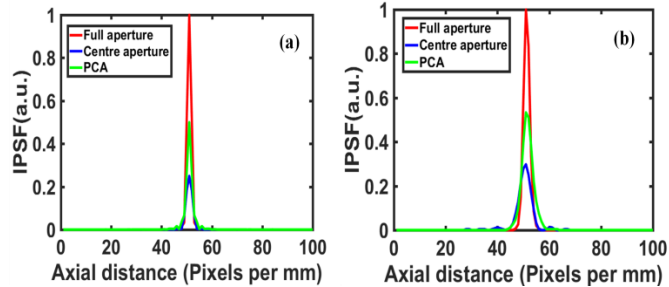

Fig. 4 IPSF comparison plots for apertures (a) simulated IPSF (b) experimentally recorded IPSF. [PCA: Polarization coded aperture].

The second PCA imaging system (a pixelated binary Fresnel zone plate (BFZP)) is experimentally designed by importing a BFZP of focal length (f=15 cm) on the SLM (pixel size=20μm).
The axial intensity of an FZP at a distance 'z' is defined as [18]

$$I(z) = \left| \sum_{n=1}^{N} (-1)^n \frac{z}{\sqrt{z^2 + n\lambda f}} \exp\left(i \frac{2\pi}{\lambda} \sqrt{z^2 + n\lambda f}\right) \right|^2 \quad (5)$$

The BFZP is polarization encoded by providing orthogonal SOPs to its alternate transparent and opaque region respectively using polarization modulation characteristics of the SLM). For this purpose, the SOP of the transparent zones is rotated by $90^0$ with respect to the SOP of the opaque (dark) zones of BFZP. This is experimentally implemented by providing grayscale '0' and '180' to the alternate opaque and transparent zones of BFZP respectively. Hence, BFZP is split into two regions of orthogonal SOPs.

Therefore, the axial intensity of polarization encoded BFZP can be given as

$$I(z)_{PCA} = I(z)_x + I(z)_y \quad (6)$$

Here, the first term is axial intensity corresponding to x-polarized zones, and the second term represents axial intensity for 90⁰ rotated (y-polarized) zones of BFZP. Since there is no interference between the orthogonal SOPs, it is expected to increase the axial intensity of PCA-enabled BFZP at its focal plane. Fig. 5 (a, b) shows the conventional BFZP and Polarization encrypted BFZP. Corresponding recorded axial intensity is demonstrated in figs. 5(c, d) respectively. It is observed that the intensity corresponding to the first-order peak is enhanced for Polarization encrypted BFZP as compared to conventional BFZP.

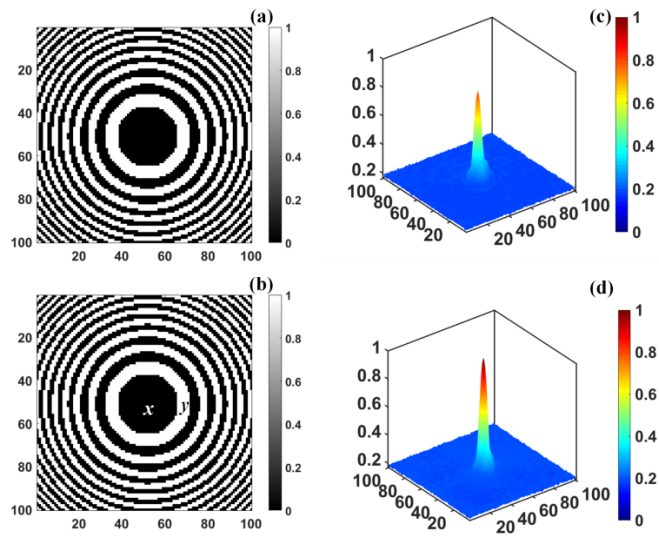

Fig.5. (a) Conventional pixelated BFZP (b) polarization coded BFZP projected on SLM (c) Recorded axial intensity at the focal plane of conventional BFZP (d) Recorded axial intensity at the focal plane of polarization coded BFZP. [BFZP: binary Fresnel zone plate].

## CONCLUSION

In conclusion, PCA based imaging system has been experimentally demonstrated to obtain improved imaging characteristics (DOF, axial intensity) of diffraction limited imaging system. The PCAs are experimentally designed using polarization modulation characteristics of an LC-SLM. The first PCA is designed using the polarization encrypted annular aperture, which yields enhanced DOF of a diffraction limited lens. Experimental results (recorded PSFs and Rayleigh range) of PCA confirm that the polarization enabled coded apertures to increase the Rayleigh range by a factor of two with minimal intensity loss as compared to conventional apertures with reduced size. In the second type of PCA, a BFZP is transformed into polarization coded BFZP by providing orthogonal SOPs of light to its alternate zones (transparent and opaque). Recorded axial intensities for BFZP allude to that the axial intensity of BFZP can be increased up to two times using the polarization coding technique. The PCA based imaging system can be utilized in order to obtain enhanced resolution without dwindling the aperture size, particularly in low illumination imaging applications. In addition, the PCA based imaging system is expected to relegate the aberration issues of diffraction limited lens as well. Immediate implications of the proposed PCA enabled imaging system involve label-free phase-contrast microscopy, correlation holography, structured light imaging applications, etc.

**Acknowledgment.** Vipin Tiwari would like to acknowledge DST-INSPIRE (IF-170861). NSB acknowledges support from DST-SERB project (YSS/2015/001894).

**Disclosures.** The authors declare no conflicts of interest.